\documentclass[10pt]{article}
\usepackage{times}
\usepackage[usenames]{color}
\usepackage{hyperref}
\def\pd#1#2{\frac{\partial#1}{\partial#2}}
\def\pdd#1#2#3{\ifx#2#3\frac{\partial^2#1}{\partial#2^2}\else\frac{\partial^2#1}{\partial#2\mkern1mu\partial#3}\fi}

\let\tint=\int
\def\eqnitem#1. {\par\medbreak\leavevmode#1$^\circ$.\enspace}
\def\Example#1.{{\sl Example\hskip.5em\relax#1\unskip.}}
\def\Remark#1.{{\sl Remark\hskip.5em\relax#1\unskip.}}
\let\BL=\biggl \let\BR=\biggr
\let\Bl=\Bigl \let\Br=\Bigr
 
\let\ds=\displaystyle
\let\ts=\textstyle
\def\Blue{\color{Blue}}

\begin{document}

\title{Equations of hydrodynamic type: exact solutions, reduction of order, transformations,
and nonlinear stability/unstability}

\author{%
A.~D.~Polyanin\footnote
{Institute for Problems in Mechanics,
  Russian Academy of Sciences,
  101~Vernadsky Avenue, bldg~1, 119526 Moscow, Russia.
  E-mail: polyanin@ipmnet.ru}
\and
S.~N.~Aristov\footnote
{Institute of Continuous Media Mechanics, Ural Branch, Russian Academy of Sciences,
  1~Acad.\ Koroleva Str., 614013 Perm, Russia.
  E-mail: asn@icmm.ru}}
\maketitle

\begin{abstract}
{\Blue Systems of hydrodynamic type equations derived from the Navier--Stokes equations
and the boundary layer equations are considered. A transformation of the Crocco
type reducing the equation order for the longitudinal velocity component is described.
The issues of nonlinear stability of the obtained solutions are studied. It is found
that a specific feature of many solutions of the Navier-Stokes equations is instability.
The nonlinear instability of solutions is proved by a new exact method, which may be
useful for the analysis of other nonlinear physical models and phenomena.}

{\Blue Self-similar, invariant, partially invariant, generalized separable and some
other exact solutions of the Navier--Stokes equations are considered, e.g., in~\hbox{[1--11]}.}
\medskip

{\Blue {\bf Keywords:} Navier--Stokes equations, boundary layer equations, Crocco transformation,
exact solutions, nonlinear instability, nonlinear stability, Calogero equation, exact methods.}
\end{abstract}

\section{Systems of hydrodynamic type equations}

{\bf Systems of equations under consideration.}
{\Blue The following systems of equation will be considered:}
\begin{eqnarray}
\pdd Ftz+F\pdd Fzz-m\Bl(\pd Fz\Br)^2&=&
\nu \frac{\partial^3F}{\partial z^3}+q(t)\pd Fz+p(t),\label{(1)}\\
\pd G t+F\pd G z-kG \pd Fz&=&\nu\pdd G zz+s(t),
\end{eqnarray}
{\Blue where $F$ is one of the components of the fluid velocity,
$G$ is an auxiliary function, and $\nu$ is the kinematic viscosity of the fluid.
For $m = 1/2$ and $m = 1$, a single equation (1) and system of equations (1), (2)
with $m= 1$ were treated in [5, 8, 10, 11], where exact solutions of
the Navier--Stokes equations and the boundary layer equations were considered.
Note that, for $m = 1$, the functions $p=p(t)$ and $q=q(t)$ in equation (1)
can be chosen arbitrarily [11]. Below, a new class of exact solutions of
the Navier--Stokes equations will be described for system (1), (2) with
$m = k = 1/2$ and $q(t) = 0$.}

{\Blue
Nonlinear equation (1) can be considered separately and equation (2) is linear
with respect to the unknown function $G$.
According to the general property of equation (1)
[8, 11], if $\widetilde F(z,t)$ is some solution to this equation,
then the function}
\begin{equation}
F=\widetilde F(z+\psi(t),t)-\psi'_t(t)
\label{(3)}
\end{equation}
{\Blue with an arbitrary function $\psi(t)$ will also be a solution to
equation (1); in addition, the function $F=-\widetilde F(-z,t)$ will also be
a solution to this equation.}
\medskip

{\bf One class of exact solutions of the Navier--Stokes equations.}
{\Blue Three-dimensional nonstationary motions of a viscous incompressible fluid are
described by the following system of the Navier-Stokes and continuity equations:}
\begin{equation}
\begin{array}{rl}
&\ds\pd {V_n}t+V_1\pd {V_n}x+V_2\pd {V_n}y+V_3\pd {V_n}z\\[12pt]
&\qquad\qquad\ds=-\frac 1{\rho}\nabla_n P
+\nu\BL(\pdd {V_n}xx+\pdd {V_n}yy+\pdd {V_n}zz\BR),\quad n=1,\,2,\,3,\\[12pt]
&\ds\pd {V_1}x+\pd {V_2}y+\pd {V_3}z=0.
\end{array}
\label{(4)}
\end{equation}
{\Blue Here $x$, $y$, and $z$ are the Cartesian coordinates;
$t$ is the time; $V_1$, $V_2$, and~$V_3$ are the fluid velocity components;
$P$ is the pressure; $\rho$ is the fluid density; and
$\nabla_1 P=\partial P/\partial x$, $\nabla_2 P=\partial P/\partial y$,
$\nabla_3 P=\partial P/\partial z$. In equations (4), it is
assumed that the mass forces are potential and included in the pressure.}

{\Blue For a viscous incompressible fluid, equations (4) admit the following
exact solutions:}
\begin{eqnarray*}
V_1&=&G-\frac12x\pd Fz,\quad V_2=-\frac12y\pd Fz,\quad V_3=F,\\[6pt]
\frac P\rho=p_0(t)&+&\frac14\alpha(t)(x^2+y^2)-s(t)x-\frac12 F^2
+\nu\pd Fz-\int \pd Ft\,dz,
\end{eqnarray*}
{\Blue where $p_0(t)$, $\alpha(t)$, and $s(t)$ are arbitrary functions of time
$t$ and the functions $F$ and $G$ depend on $z$ and $t$ and satisfy the equations}
\begin{eqnarray}
\pdd Ftz+F\pdd Fzz-\frac12\Bl(\pd Fz\Br)^2&=&
\nu \frac{\partial^3F}{\partial z^3}+\alpha(t),\label{(5)}\\
\pd Gt+F\pd Gz-\frac12G\pd Fz&=&\nu\pdd Gzz+s(t).
\end{eqnarray}
{\Blue System of equations (5), (6) is a special case of system (1), (2) with
$m=k=1/2$ and $q(t)=0$.}

\section{Reduction of order of equation (1) and its generalizations with the use
the Crocco type transformation}

{\bf Reduction of order of equation (1).}
{\Blue Let us introduce the notation}
\begin{equation}
\eta=\pd Fz,\quad \ \Phi=\pdd Fzz.
\label{(7)}
\end{equation}

{\Blue Transferring the term $mF_z^2$ (here and below, the brief
record of derivatives is used) to the right-hand side of equation (1),
dividing both the resulting equation by $F_{zz}=\Phi$, differentiating with respect
to~$z$, and taking into account formulas (7),
we eventually obtain the following equation:}
\begin{equation}
\frac{\Phi_t}\Phi-\frac{F_{zt}\Phi_z}{\Phi^2}+\eta=\pd{}z\frac {\nu \Phi_{z}+m\eta^2+q(t)\eta+p(t)}\Phi.
\label{(8)}
\end{equation}

{\Blue Now let us pass in equation (8) from the old variables $t$, $x$, $F=F(x,t)$
to the new variables $t$,~$\eta$,~$\Phi=\Phi(t,\eta)$,
where $\eta$ and $\Phi$ are defined by formulas (7).
The derivatives are transformed as follows:}
\begin{eqnarray*}
\pd{}z&=&\pd\eta z\pd{}\eta=F_{zz}\pd{}\eta=\Phi\pd{}\eta,\\[6pt]
\pd{}t&=&\pd{}t+\pd\eta t\pd{}\eta=\pd{}t+F_{zt}\pd{}\eta.
\end{eqnarray*}
{\Blue As a result, equation (8) reduces to the second-order equation}
$$
\frac{\eta}\Phi-\pd{} t\frac 1\Phi=
\pd{}\eta\BL(\frac {m\eta^2+q\eta+p}\Phi\BR)+\nu\pdd\Phi\eta\eta
$$
{\Blue which can be rewritten as follows:}
\begin{equation}
\pd\Phi t+(m\eta^2+q\eta+p)\pd\Phi\eta=[(2m-1)\eta+q]\Phi+\nu\Phi^2\pdd\Phi\eta\eta.
\label{(9)}
\end{equation}
{\Blue Here and below, for brevity, the arguments of the functions
$p(t)$, $q(t)$, and $s(t)$ will often be omitted.}

{\Blue Note that, in the degenerate case (inviscid fluid, $\nu=0$), the initial
nonlinear second-order equation (1) is reduced to the first-order linear equation (9),
which can be completely integrated using the method of characteristics.}

{\Blue If a solution of the initial equation (1) is known, then formulas (7) define a solution
to equation (9) in the parametric form.}

{\Blue Let $\Phi=\Phi(\eta,t)$ be some solution to equation (9). Then, the corresponding
solution to the initial equation (1) can also be presented in parametric form
as follows:}
$$
z=\int\frac{ds}{\Phi(s,t)}+\psi(t),\quad F=\int\frac{s\,ds}{\Phi(s,t)}-\psi'_t(t),
$$
{\Blue where $\psi(t)$ is an arbitrary function (in the integrals, $t$ is treated as a parameter).}

\medbreak
{\bf Transformation of system (1), (2).}
{\Blue Assuming that $F_{zz}\not=0$, let us pass in system (1), (2) from the old
variables $t$, $z$, $F$ to the new variables $t$, $\eta$, $\Phi$ according to
formulas (7). Then, equation (1) is transformed into equation (9) and equation (2)
is transformed into the following equation:}
\begin{equation}
\pd G t+(m\eta^2+q\eta+p)\pd G \eta-k\eta G =\nu\Phi^2\pdd G \eta\eta+s(t).
\label{(10)}
\end{equation}
{\Blue In deriving this equation, we used a representation for the mixed derivative
obtained from equation (1).}

{\Blue At $m = k$, equation (10) has exact solutions of the following form:}
\begin{equation}
G=A\eta + B\Phi+C,
\label{(11)}
\end{equation}
{\Blue where $A=A(t)$, $B=B(t)$, and $C=C(t)$
are unknown functions determined by an appropriate system of ordinary
differential equations (at $m=k\not=1$, we have $B=0$). To prove this fact,
one should substitute expression (11) into equation (10) and take into
account equation (9). Formula (11) will be used below for the representation
of solutions to equation (2) via solutions to equation (1).}
\medskip

{\bf Some generalizations.}
{\Blue Let us consider the nonlinear $n$th-order equation}
\begin{equation}
\pdd Fzt+[a(t)F+b(t)z]\pdd Fzz=H\Bl(t,\pd Fz,\pdd Fzz, \pd{^3F}{z^3},\dots, \pd{^nF}{z^n}\Br),
\label{(12)}
\end{equation}
{\Blue which generalize equation (1) and allow for order reduction.
Passing in equation (12) from $t$, $x$, $F=F(x,t)$ to the new variables
$t$,~$\eta$,~$\Phi=\Phi(t,\eta)$, where $\eta$ and $\Phi$ are defined by
formulas (7), we obtain the $(n - l)$th-order equation}
\begin{equation}
\frac{a(t)\eta+b(t)}\Phi-\pd{}t\frac 1\Phi=\pd{}\eta
\BL[\frac 1\Phi H\Bl(t,\eta,\Phi,\Phi\pd\Phi\eta,\dots,\pd{^{n-2}\Phi}{z^{n-2}}\Br)\BR],
\label{(13)}
\end{equation}
{\Blue in which the high-order derivatives are calculated using the formulas}
$$
\pd{^{k} F}{z^{k}}=\pd{^{k-2}\Phi}{z^{k-2}}=\Phi\pd{}\eta\pd{^{k-3}\Phi}{z^{k-3}},\quad
\pd{}z=\Phi\pd{}\eta,\quad k=3,\dots,n.
$$

{\Blue In the special case of a second-order equation with
$n=1$, $a(t)=-1$, $b(t)=0$, and $H=H(F_z)$, equation (12) reduces to the
Calogero equation, which was considered in [8, 12, 13].
It is evident that a more general equation (12) with $n=1$, $H=H(t,F_z)$,
and arbitrary functions $a(t)$ and $b(t)$
(it is logical to call this the generalized Calogero equation),
can be reduced using the Crocco type transformation to the first-order equation (13),
which becomes linear after the substitution $\Phi=1/\Psi$.}

\section{Representation of solutions to equation (2) via solutions to equation (1)}

{\bf The case of \boldmath $m=k=1$.}
{\Blue Let $F=F(z,t)$ be a solution to equation (1).
Then, by virtue of relations (11) and (7), equation (2) has the solution}
\begin{equation}
 G=A'_t+Aq+A\pd Fz+B\pdd Fzz,
\label{(14)}
\end{equation}
{\Blue where the functions $A = A(t)$ and $B= B(t)$ satisfy the ordinary
differential equations}
\begin{eqnarray}
A''_{tt}+qA'_t+(p+q'_t)A&=&s,\label{(15)}\\
B'_t+qB&=&0.
\end{eqnarray}
{\Blue The proof is carried out by eliminating the function $G$ from (2) and (14),
followed by the comparison of the obtained expression to both equation (1)
and the equation resulting from the differentiation of equation (1) with respect to~$z$.}

{\Blue The general solution to equation (16) is as follows:
$\ds B=C\exp\Bl(-\tint q\,dt\Br)$, where $C$ is an arbitrary constant.}
\medskip

{\bf\boldmath The case of $m=k\not=1$.}
{\Blue In this case, by virtue of relations (11) and (7), equation (2) has the solution}
\begin{equation}
 G=\frac 1m(A'_t+Aq)+A\pd Fz,
\label{(17)}
\end{equation}
{\Blue where function $A=A(t)$ satisfies the ordinary differential equation}
\begin{equation}
A''_{tt}+qA'_t+(mp+q'_t)A=ms.
\label{(18)}
\end{equation}

{\Blue The exact solutions to equation (1) can be found in [5, 8, 10, 11].
Formulas (14)-(18) are used to obtain the corresponding exact solutions to equation (2).}

\section{Use of linear transformations for constructing exact solutions to equations (1)}

{\Blue Using the following linear transformation with respect to the unknown function,}
\begin{equation}
\begin{array}{rl}
F&=a(t)f(\tau,\xi)+b(t)z+c(t),\\[6pt]
\xi&=\ds\lambda(t)z+\sigma(t),\quad \tau=\int \lambda^2(t)\,dt+C_0,
\end{array}
\label{(19)}
\end{equation}
{\Blue where $a=a(t)$, $b=b(t)$, $c=c(t)$, $\lambda=\lambda(t)$, and
$\sigma=\sigma(t)$ are arbitrary functions, we write equation (1)
in the following form:}
\begin{equation}
\pdd f\tau\xi+[\widetilde a(\tau)f+\widetilde b(\tau)\xi+\widetilde c(\tau)]\pdd f\xi\xi-m\widetilde a(\tau)\Bl(\pd f\xi\Br)^2=
\nu \frac{\partial^3f}{\partial \xi^3}+\widetilde q(\tau)\pd f\xi+\widetilde p(\tau),
\label{(20)}
\end{equation}
{\Blue where}
\begin{equation}
\begin{array}{rl}
\widetilde a&=\ds\frac a\lambda, \quad \ \widetilde b=\frac 1{\lambda^3}(b\lambda+\lambda'_t),\quad
\widetilde c=\frac 1{\lambda^3}(c\lambda^2-b\lambda\sigma+\lambda\sigma'_t-\sigma\lambda'_t),\\[12pt]
\widetilde q&=\ds\frac 1{a\lambda^3}[aq\lambda+2mab\lambda-(a\lambda)'_t],\quad
\widetilde p=\frac 1{a\lambda^3}(p+bq+mb^2-b'_t).
\end{array}
\label{(21)}
\end{equation}
{\Blue The argument of the functions on the left-hand side of the equations is $\tau$
and that on the right-hand side is $t$; the variables $\tau$ and $t$ are linked by
the last relation in (19).}

{\Blue The presence of a large number (from five to seven) of arbitrary functions in
equations (1) and (19) allows us to construct various exact solutions to equation (1).}
\medskip

{\Blue \Example 1.
Assuming sequentially that (20)
$f=\xi^{-1}$, $f=Ae^\xi+Be^{-\xi}$,
$f=A\sin\xi+B\cos\xi$, and $f=\tanh\xi$ at $m=1$ in
transformation (19) and equation (20), we can obtain the solutions given in [8, 11]
by selecting appropriate arbitrary functions. Equation (20) is also satisfied
by the functions $f=(1\pm e^{\xi})^{-1}$ and $f=\tan(\xi+A)$, which give new solutions.}
\medskip

{\Blue \Example 2. Assuming that}
\begin{equation}
\widetilde a=C_1,\quad \widetilde b=C_2,\quad \widetilde c=C_3,\quad
\widetilde q=C_4,\quad \widetilde p=C_5,
\label{(22)}
\end{equation}
{\Blue where $C_n$ are arbitrary constants,
we obtain an ordinary differential equation for the function $f=f(\xi)$ from
equation (20). In this case, relations (21) under conditions (22) represent
a system of ordinary differential equations for the functional coefficients
of transformation (19). In equation (1) with $m = 1$ and in transformation (19)
with constraints (21) and (22), two functions can be set arbitrarily
(it should be recalled that $p=p(t)$ and $q=q(t)$ are arbitrary functions).
Stationary solutions $f=f(\xi)$ of equation (20) generate
nonstationary traveling-wave solutions (19) of the initial equation~(1).}
\medskip

{\Blue \Example 3. Now we assume that}
\begin{equation}
\begin{array}{rl}
\widetilde a&=C_1\tau^{\ts-\frac{k+1}2},\quad \widetilde b=C_2\tau^{-1},\quad
\widetilde c=C_3\tau^{\ts-\frac12},\\[6pt]
\widetilde q&=C_4\tau^{-1},\quad \widetilde p=C_5\tau^{\ts\frac{k-3}2},\quad
\ds\tau=\int \lambda^2(t)\,dt+C_0,
\end{array}
\label{(23)}
\end{equation}
{\Blue where $k$ and $C_n$ are arbitrary constants.
In this case, equation (20) admits self-similar solutions of the following form:}
$$
f=\tau^{k/2}h(\zeta),\quad \zeta=\xi\tau^{-1/2},
$$
{\Blue where the function $h=h(\zeta)$ satisfies the ordinary differential equation}
\begin{equation}
\Bl(\frac{k-1}2-C_4\Br)h'_\zeta
+\Bl[C_1h+\Bl(C_2-\frac12\Br)\zeta+C_3\Br]h''_{\zeta\zeta}-mC_1(h'_\zeta)^2=
\nu h'''_{\zeta\zeta\zeta}+C_5.
\label{(24)}
\end{equation}

{\Blue Substituting expressions (23) into (21), we obtain a system of integro-differential
equations for the functional parameters of the initial equation (1) and
transformation (19). Note that, using the substitution $\lambda=\sqrt{\varphi'_t}$,
and taking into account the relation $\tau=\varphi(t)+C_0$, we arrive at
a standard system of ordinary differential equations.
As a result, a non-self-similar solution of the form (19) is obtained.}

\section{Nonlinear analysis of stability/instability of solutions}

{\bf Analysis of stability/instability of solutions based on equation (2).}
{\Blue Consider system (1), (2) with $m=k=1$ and $s=0$ obtained in [11].
To analyze stability/instability of solutions, we use formula (14) and
equations (15) and (16) relating the solutions of system (1), (2).
It is important to note that, in many cases, there is no necessity to know
the explicit form of the function $F$.}

\penalty -50
{\Blue First, let us study problems with a stationary longitudinal velocity,
which correspond to the case of $F=F(z)$, $p={\rm const}$, and $q={\rm const}$.
In this case, the solution to equation (15) depends on the sign of the
discriminant $\Delta=q^2-4p$:}
\begin{equation}
A(t)=\cases{\ds \exp\Bl(-\frac{qt}2\Br)\Bl[C_1\exp\Bl(\frac{t\sqrt{\Delta}}2\Br)+C_2\exp\Bl(-\frac{t\sqrt{\Delta}}2\Bl)\Br] &
if \ $\Delta>0$,\cr
\ds \exp\Bl(-\frac{qt}2\Br)\Bl[C_1\sin\Bl(\frac{t\sqrt{|\Delta|}}2\Br)+C_2\cos\Bl(\frac{t\sqrt{|\Delta|}}2\Br)\Br] &
if \ $\Delta<0$,\cr
\ds \exp\Bl(-\frac{qt}2\Br)(C_1t+C_2) & if \ $\Delta=0$,\cr}
\label{(25)}
\end{equation}
{\Blue where $C_1$ and $C_2$ are arbitrary constants.
Further, for the sake of simplicity, we assume $B=0$ in equations (14) and (16).
In the analysis, we consider the following two cases.}

{\Blue \eqnitem 1. {\it Nondegenerate case \/} $F_z\not\equiv 0$.
For $q < 0$ (and arbitrary $p$) or $p < 0$ (and arbitrary $q$),
solutions (14) and (25) (with $C_1\not=0$) increase exponentially as $t\to\infty$.
Therefore, the specified values of the parameters $p$ and $q$ determine
the domain of nonlinear instability of system (1), (2) for any limited
stationary profile of the longitudinal component of the velocity $F(z)$
(other than a constant). The point $p=q=0$ also belongs to the domain of
instability of system (1), (2).}

{\Blue Indeed, by choosing suitable values of the constants $C_1$ and $C_2$ and using
equations (14) and (25), we can make the initial value of $|G|_{t=0}$
(interpreted as the initial perturbation relative to the trivial solution $G=0$
of equation (2)) smaller than any preset $\varepsilon$. However,
for $q<0$ (and arbitrary $p$) or $p < 0$ (and arbitrary $q$),
we have $|G|\to\infty$ as $t\to \infty$.
This means that arbitrary small perturbations of the solutions to system (1), (2)
exhibit unbounded growth with time.}

{\Blue \Remark.
If $F\to F_1$ at $z\to -\infty$ and
$F\to F_2$ as $z\to +\infty$ ($F_1,F_2={\rm const}$),
then solution (14) at $A=0$, $B\not=0$ tends to zero as $z\to\pm\infty$.}
\medskip

{\Blue For $q=0$ and $p>0$, solution (25) and hence solution (14) are periodic.
The inequalities $q\ge 0$, $p\ge 0$ ($|p|+|q|\not=0$)
determine the domain of conditional stability of the solutions under consideration.}

{\Blue It is important to emphasize that (i) here we are speaking of the {\it nonlinear
instability\/}; (ii) all the results and solutions obtained above are exact
(rather than linearized, which is the case in the theory of linear stability);
and (iii) various assumptions, expansions, and approximations inherent in many
nonlinear theories are not used either [2, 14, 15]).}
\medskip

{\Blue \Example 1. The stationary spatially periodic solution}
$$
F=a\sin(\sigma z+b),\quad G=0\quad \ (p=-a^2\sigma^2,\quad q=\nu\sigma^2)
$$
{\Blue of system (1), (2) is unstable for any values of  $a$, $b$, and $\sigma$
($a\not=0$, $\sigma\not=0$).}
\medskip

{\Blue \Example 2. The stationary monotonic restricted solution}
$$
F=-6\nu \sigma \tanh(\sigma z+b),\quad G=0\quad \ (p=0,\quad q=8\nu \sigma ^2)
$$
{\Blue of system (1), (2) is stable.}

\medbreak
{\Blue All conclusions concerning the stability/instability of solutions presented above,
as well as formulas (14) and (25), remain valid for any nonstationary solutions
$F=F(z,t)$, $G=G(z,t)$ (under the condition that the derivative $F_z\not\equiv 0$
is bounded) of system (1), (2) with $p={\rm const}$ and $q = {\rm const}$.}

{\Blue According to the above considerations, three quarters of the plane of
parameters $p$, $q$ correspond to non-stationary solutions.
It is important to note that the flow instability described above is not
associated with a specific velocity profile and is realized due to equation (2)
governing the transverse components of the fluid velocity. Since the fluid
viscosity $\nu$ does not enter equations (15) and formulas (25),
the above results do not depend on the Reynolds number; i.e.,
the instability of solutions takes place not only at large but also at small
Reynolds numbers ($0 < {\rm Re} < \infty$).}
\medskip

{\Blue \Remark.
Likewise, we can use equations (14) and (15) to study instability of the
nonstationary solutions of system (1), (2) with variables $p=p(t)$ and $q=q(t)$.}

{\Blue \eqnitem 2.
{\it Degenerate case\/} $F_z\equiv 0$. Let}
\begin{equation}
F=a={\rm const}\qquad (\hbox{if \ $p=0$}).
\label{(26)}
\end{equation}
{\Blue Then any solution to equation (2) that, when passing from $t$, $z$
to the new variables $t$, $\xi=z-at$, reduces to the classical heat equation,
is stable for any values of the parameters $a$ and $q$.}
\medskip

{\bf\boldmath Analysis of stability of solution $F={\rm const}$ of equation (l) for
$p=0$.}
{\Blue \eqnitem 1. Let us study stability of the trivial solution
$F=0$ of equation (1) for $p=0$ and various values of the parameter
$q={\rm const}$. Equation (1) admits the following exact solution:}
\begin{equation}
F=\varepsilon e^{ikz+\lambda t},\quad \ \lambda=q-\nu k^2,
\label{(27)}
\end{equation}
{\Blue where $\varepsilon$, $k$, and $\lambda$
are real quantities. This solution is also a solution to the linearized equation (1)
with the quadratic terms discarded.
The absolute value of the difference between solution (27) and the trivial solution
at the initial moment of time is $|\varepsilon|$
(this difference can be made arbitrarily small by choosing a suitable $\varepsilon$
value).}

{\Blue For $q-\nu k^2>0$, the trivial solution will be unstable,
whereas it will be stable for $q-\nu k^2<0$.
The stability boundary is the parabola $q=\nu k^2$ on the $k$,~$q$ plane. For
decreasing fluid viscosity, $\nu\to 0$ (which corresponds to increasing
Reynolds numbers), the branches of this parabola tend to the
line $q=0$ and the domain of instability expands to become, in the limiting case,
the entire upper half-plane $q>0$. Increasing $\nu$ or $k$ results in the expansion
of the stability domain.
Since the parameter $k$ can be set arbitrarily, then, for any $q>0$ we can
achieve instability of the trivial solution by choosing an appropriate $k$ value.}

{\Blue \eqnitem 2. Consider an arbitrary stationary solution of the form (26).
Instead of solution (27), we use the following function:}
\begin{equation}
F=\varepsilon e^{ik(z-at)+\lambda t}+a,\quad \ \lambda=q-\nu k^2,
\label{(28)}
\end{equation}
{\Blue Owing to property (3) with $\psi(t)=-at$, this function is also a solution
to equation (1). The absolute value of the difference between equations (26)
and (28) at the initial time can be made arbitrarily small by choosing a
suitable $\varepsilon$ value. All criteria of stability and instability of
solution (26) depending on the parameters $k$ and $q$ remain the same as those for
the trivial solution.}
\medskip

{\Blue \Remark.
It follows from the above results that, for $p = 0$ and $q\le 0$, only a constant
profile of the longitudinal velocity component ($F= {\rm const}$) is stable.}

\section{Transformation and exact solutions of the boundary layer equations}

{\Blue Nonstationary equations of the plane boundary layer in terms of the stream
function $w$ are reduced to a single third-order equation [8]:}
$$
\pdd wty+\pd wy\pdd wxy-\pd wx\pdd wyy=\nu\pd{^3w}{y^3}+f(x,t).
$$

{\Blue Let us pass from $t$, $x$, $y$, and $w$ to the new
variables $t$, $x$, $\eta$, and $\Phi$, where $\eta$, $\Phi$ are the generalized Crocco
variables defined as follows:}
$$
\eta=\pd wy,\quad \ \Phi(x,t,\eta)=\pdd wyy.
$$
{\Blue As a result, we obtain the following second-order equation,}
$$
\pd\Phi t+\eta \pd\Phi x+f(x,t)\pd\Phi\eta=\nu\Phi^2\pdd\Phi\eta\eta,
$$
{\Blue which is reduced by the substitution $\Phi=1/\Psi$ to the nonlinear heat equation:}
$$
\pd\Psi t+\eta \pd\Psi x+f(x,t)\pd\Psi\eta=\nu\pd{}\eta\Bl(\frac 1{\Psi^2}\pd\Psi\eta\Br).
$$

{\Blue \eqnitem 1.
Let us first consider the special case of $f(x,t)=f(t)$ and for special
solutions of the following form:}
$$
\Psi=Z(\xi,\tau),\quad \xi=x-\eta t+\tint tf(t)\,dt,\quad \tau=\frac 13t^3.
$$
{\Blue Here, we have the solvable equation}
\begin{equation}
\pd Z\tau=\nu\pd{}\xi\Bl(\frac 1{Z^2}\pd Z\xi\Br),
\label{(29)}
\end{equation}
{\Blue which can be reduced to the linear heat equation [8].}

{\Blue \eqnitem 2.
Consider a more general case of $f(x,t)=f(t)x+g(t)$
and seek the following special solutions:}
$$
\Psi=Z(\xi,\tau),\quad \xi=\varphi(t)x+\psi(t)\eta+\theta(t),\quad \tau=
\tint \psi^2(t)\,dt,
$$
{\Blue where the functions $\varphi=\varphi(t)$, $\psi=\psi(t)$, and $\theta=\theta(t)$
are determined by the following linear system of ordinary differential equations}
$$
\varphi'_t+f\psi=0,\quad
\psi'_t+\varphi=0,\quad
\theta'_t+g\psi=0.
$$
{\Blue As a result, we also arrive at the solvable equation (29).}

\section*{Acknowledgments}

{\Blue The work was carried out under partial financial support of the Russian
Foundation for Basic Research (grants  No.~08-01-00553, No.~08-08-00530,
\hbox{No.~07-01-96003-r$_{-}$ural$_{-}$a}, and No.~09-01-00343).}

\end{document}